# Structural, Superconducting and Magnetic Properties of $La_{3-x}R_xNi_2B_2N_{3-\delta}$ ($R$ = Ce, Pr, Nd)

T. Ali, E. Bauer, G. Hilscher, H. Michor

Institute of Solid State Physics, Vienna University of Technology, Vienna A-1040, Austria

**Abstract.** We report on structural and superconducting properties of $La_{3-x}R_xNi_2B_2N_{3-\delta}$ where La is substituted by the magnetic rare-earth elements Ce, Pr, Nd. The compounds $Pr_3Ni_2B_2N_{3-\delta}$ and $Nd_3Ni_2B_2N_{3-\delta}$ are characterized for the first time. Powder X-ray diffraction confirmed all samples $R_3Ni_2B_2N_{3-\delta}$ with $R$ = La, Ce, Pr, Nd and their solid solutions to crystallize in the body centered tetragonal $La_3Ni_2B_2N_3$ structure type. Superconducting and magnetic properties of $La_{3-x}R_xNi_2B_2N_{3-\delta}$ were studied by resistivity, specific heat and susceptibility measurements. While $La_3Ni_2B_2N_{3-\delta}$ has a superconducting transition temperature $T_c \sim 14$ K, substitution of La by Ce, Pr, and Nd leads to magnetic pair breaking and, thus, to a gradual suppression of superconductivity. $Pr_3Ni_2B_2N_{3-\delta}$ exibits no long range magnetic order down to 2 K, $Nd_3Ni_2B_2N_{3-\delta}$ shows ferrimagnetic ordering below $T_C$ =17 K and a spin reorientation transition to a nearly antiferromagnetic state at 10 K.

## Introduction

Rare-earth nickel borocarbide compounds $RNi_2B_2C$ ($R$ = rare earth) revealed a rich interplay of magnetism and superconductivity with relatively high superconducting transitions $T_c$ up to 16 K and antiferromagnetic ordering $T_N$ up to 20 K (see e.g. Refs. [1,2]). A closely related quaternary rare earth nickel boronitride system, $La_3Ni_2B_2N_{3-\delta}$, with rock salt type LaN triple layer sheets in between NiB layers shows superconductivity with $T_c \sim 12$ - 15 K [3,4] similar to the borocarbides, however, substitutions of La by magnetic rare earth elements have not yet been reported except for a cerium based homologue, $Ce_3Ni_2B_2N_{3-\delta}$. The latter was synthesized as powder material via a metathesis reaction by Glaser et al. [5]. Magnetic susceptibility measurements down to 5 K revealed neither superconductivity nor magnetic ordering. In this paper we report on the preparation of novel bulk metallic samples $La_{3-x}R_xNi_2B_2N_{3-\delta}$ with $R$ = Pr and Nd which are studied by means of X-ray diffraction, susceptibility, specific heat and transport measurements.

## Experimental details

Polycrystalline samples of $La_{3-x}R_xNi_2B_2N_{3-\delta}$ were prepared by inductive levitation melting of $(La,R)_3Ni_2B_2$ precursor alloys in $Ar/N_2$ atmosphere such that the N-stoichiometry is slowly increased. The N-adsorption is determined by two independent ways: by measuring the mass gain after each melting cycle and by measuring the pressure drop within the recipient. The combination of these methods allows to determine the nominal nitrogen stoichiometry of $La_{3-x}R_xNi_2B_2N_{3-\delta}$ within 0.05 formula units when preparing bulk samples of about 10 g. The samples were finally annealed in a vacuum furnace at 1100° C for 1 week. After this initial heat treatment smaller pieces were sealed in quartz ampoules for additional heat treatments with final quenching of the ampoules in water. The phase purity of the samples was checked by powder X-ray diffraction (XRD) on a Siemens D5000 diffractometer using monochromated Cu-$K_\alpha$ radiation. Full profile Rietveld refinements were carried out using the FULLPROF program [6]. Resistivity measurements were performed on bar shaped samples with a standard four probe method. Magnetisation and dc susceptibility measurements were carried out on a 6T Cryogenic SQUID magnetometer. Specific heat measurements were performed on 2-3 g samples employing an adiabatic step heating technique.

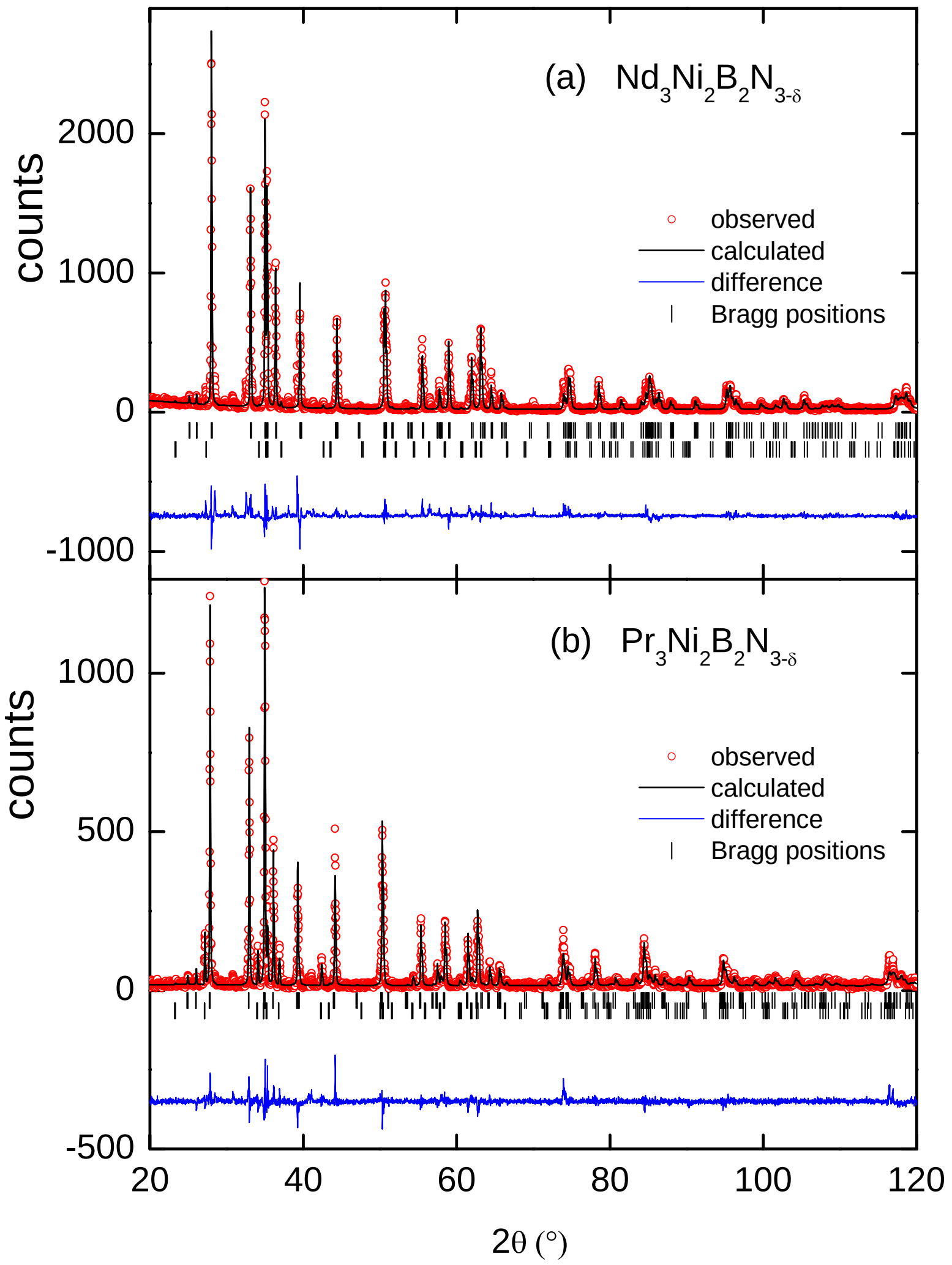

Fig. 1. Room temperature XRD pattern of $Nd_3Ni_2B_2N_{3-\delta}$ and $Pr_3Ni_2B_2N_{3-\delta}$; Bragg reflections are indicated for $La_3Ni_2B_2N_3$ (upper bars) and LaNiBN (lower bars) structure types; solid lines derive from the Rietveld refinement of these two phases.

## Results and discussion

Powder XRD results of all samples $La_{3-x}R_xNi_2B_2N_{3-\delta}$ ($R$ = Ce, Pr, Nd) display the body centered tetragonal $La_3Ni_2B_2N_3$ structure-type (space group *I4/mmm*). Minor admixtures of the related two-layer boronitride (La,*R*)NiBN are observed with typical phase fractions of about 10%. Two exemplary powder XRD patterns of $Nd_3Ni_2B_2N_{3-\delta}$ and $Pr_3Ni_2B_2N_{3-\delta}$ and their Rietveld refinements performed with the FULLPROF suite are shown in Fig. 1a,b. In the case of $Pr_3Ni_2B_2N_{3-\delta}$ we identify the two-layer variant PrNiBN as the main impurity phase with a refined fraction of about 10 %, whereas in $Nd_3Ni_2B_2N_{3-\delta}$ the two-layer phase is less significant (about 5%), but some reflections of unidentified phases are visible in the pattern.

| | a [nm] | c [nm] | V [nm$^3$] |
|---|---|---|---|
| $La_3Ni_2B_2N_{3-\delta}$ | 0.372 | 2.052 | 0.284 |
| $Ce_3Ni_2B_2N_{3-\delta}$ | 0.357 | 2.025 | 0.259 |
| $Pr_3Ni_2B_2N_{3-\delta}$ | 0.362 | 2.051 | 0.272 |
| $Nd_3Ni_2B_2N_{3-\delta}$ | 0.360 | 2.049 | 0.265 |

Table 1: Tetragonal lattice parameters *a*, *c*, and the unit cell volume *V* of $R_3Ni_2B_2N_{3-\delta}$.

The structure parameters of the series $R_3Ni_2B_2N_{3-\delta}$ summarized in Table 1 reveal a volume reduction which tends to follow the lanthanide contraction, however, with a significantly larger contraction in the case of $Ce_3Ni_2B_2N_{3-\delta}$ which is attributed to intermediate valence of Ce-ions, where cerium adopts a smaller ionic radius than that expected for $Ce^{3+}$. A more detailed discussion of structural and intermediate valence features of $Ce_3Ni_2B_2N_{3-\delta}$ and of the solid solution $(La,Ce)_3Ni_2B_2N_{3-\delta}$ are reported elsewhere [7].

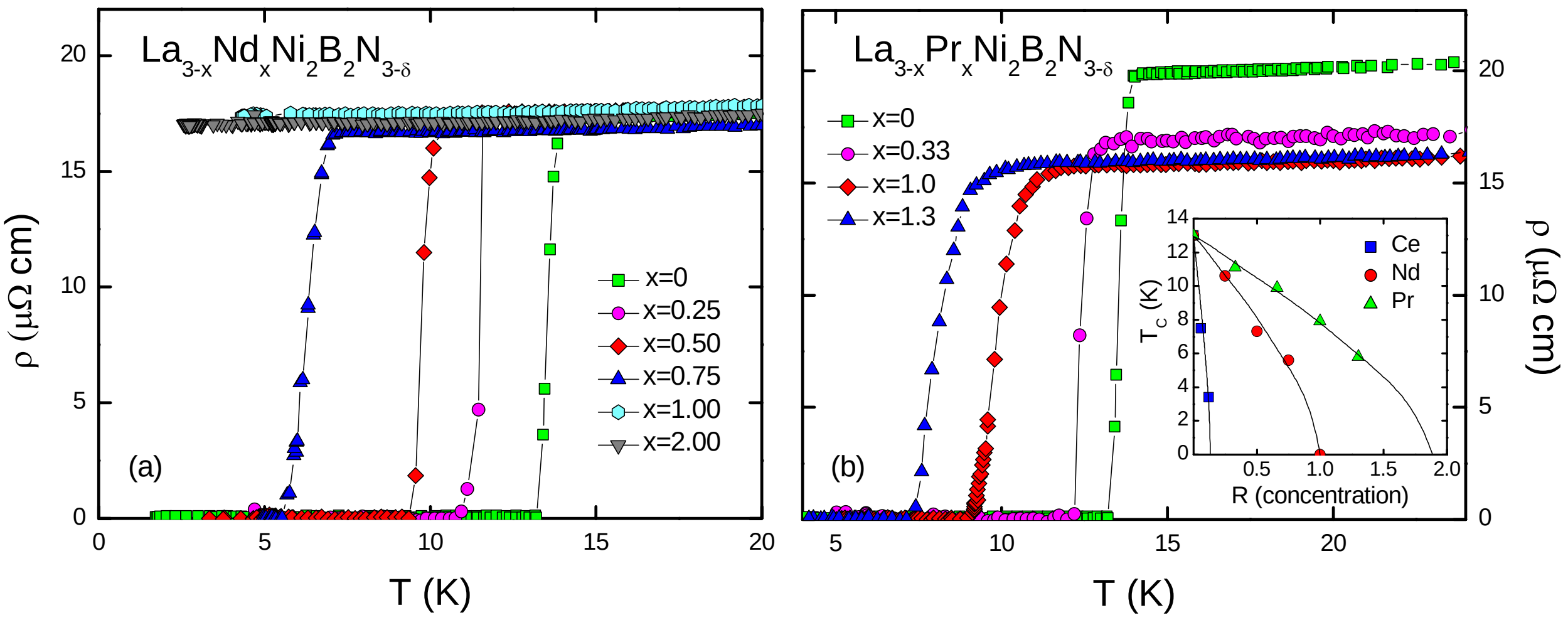

Fig. 2. Resistivity measurements of solid solutions $La_{3-x}Nd_xNi_2B_2N_{3-\delta}$ (a) and $La_{3-x}Pr_xNi_2B_2N_{3-\delta}$ (b); inset in (b): superconducting transition temperatures $T_c$ versus *R*-concentration x; solid lines are fits in terms of the Abrikosov-Gor'kov pair breaking theory.

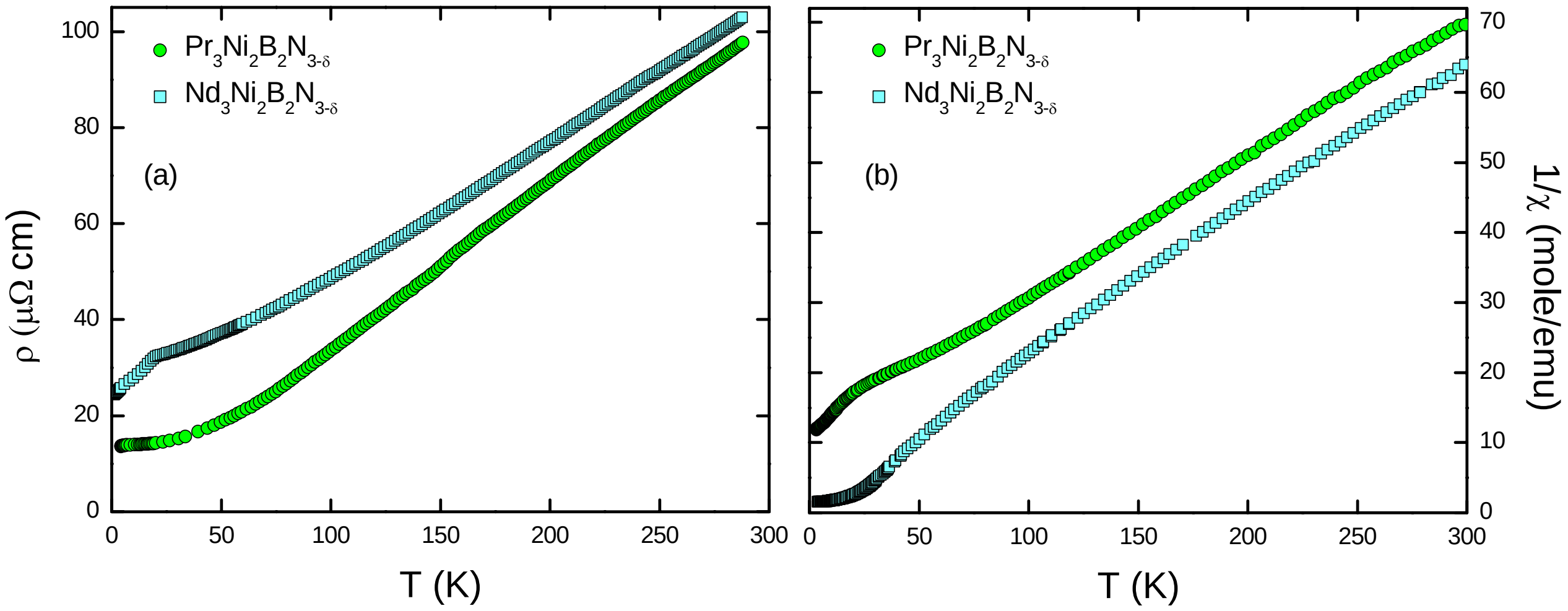

Fig. 3. Temperature dependent electrical resistivity (a) and inverse susceptibility of $Pr_3Ni_2B_2N_{3-\delta}$ and $Nd_3Ni_2B_2N_{3-\delta}$ measured at a static magnetic field of 3 T (b).

The antagonistic nature of superconductivity and magnetism is well known from earliest studies on the effect of paramagnetic rare-earth impurities in La metal [8] and has been explained theoretically by Abrikosov and Gor'kov (AG) [9] as the consequence of the exchange interaction between the localized magnetic moments and the conduction electrons which leads to pair breaking. Investigating solid solutions $La_{3-x}R_xNi_2B_2N_{3-\delta}$ with $R$ = Ce [7], Pr, Nd by means of susceptibility and resistivity measurements (see Fig. 2a,b) reveals the expected gradual suppression of superconductivity due to AG-type magnetic pair breaking (see inset of Fig. 2b) with critical $R$ concentrations $x_{crit.}$ of about 0.15, 1.9 and 1.0 for Ce, Pr, and Nd, respectively. Thereby, the pair breaking strength (i.e. suppression rate of $T_c$ or $1/x_{crit.}$) of Pr and Nd scale roughly with the deGennes factor dG = $(g_J-1)^2 J(J+1)$ where $g_J$ is the Landé factor and $J$ the total angular momentum of the magnetic rare earth ion ($dG^{Pr}$ = 0.8 and $dG^{Nd}$ = 1.84). Despite of the smaller dG = 0.18 of Ce, we observe a significantly stronger pair-breaking effect in the solid solution $La_{3-x}Ce_xNi_2B_2N_{3-\delta}$ which we attribute to the intermediate valence behaviour of Ce [7].

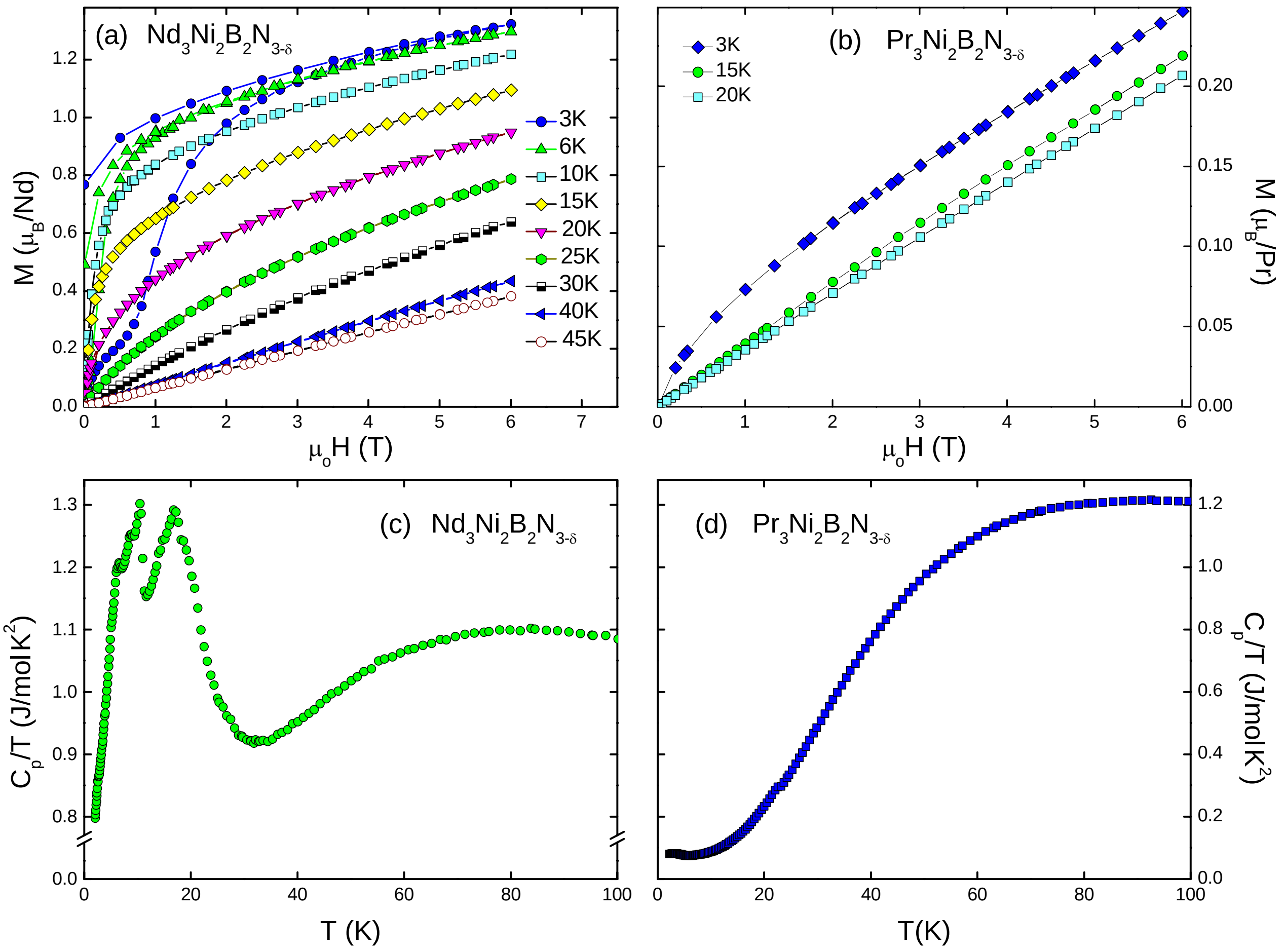

Fig. 4. Isothermal magnetisation *M(H)* measured at temperatures as labeled (a,b) and zero-field specific heat displayed as *C/T* vs. *T* (c,d) of $Nd_3Ni_2B_2N_{3-\delta}$ and $Pr_3Ni_2B_2N_{3-\delta}$, respectively.

The magnetic properties of the novel compounds $Pr_3Ni_2B_2N_{3-\delta}$ and $Nd_3Ni_2B_2N_{3-\delta}$ were studied by means of electrical resistivity, magnetic susceptibility, isothermal magnetisation and specific heat measurements. The zero-field resistivity data displayed in Fig. 3a reveal normal conducting metallic behaviour for both compounds. The resistivity of $Nd_3Ni_2B_2N_{3-\delta}$ exhibits a sharp kink near 20 K which indicates a magnetic phase transition, whereas the $Pr_3Ni_2B_2N_{3-\delta}$ data do not exhibit any anomalies. We have analysed the inverse static susceptibility measured at an applied magnetic field of 3 T (see Fig. 3b) of $Pr_3Ni_2B_2N_{3-\delta}$ and $Nd_3Ni_2B_2N_{3-\delta}$ for temperatures $T > 60$ K in terms of the Curie-Weiss model, $\chi = C/(T-\theta_p)$ with $\theta_p$ being the paramagnetic Curie temperature and C the Curie constant, $C = \mu_0 N_A \mu_{eff}^2/3k_B$. The latter allows to calculate the effective paramagnetic moment $\mu_{eff}$ yielding 3.1 $\mu_B$/Pr and 3.2 $\mu_B$/Nd. Due to crystal field effects, these numbers are slightly reduced as compared to the free ion values of $\mu_{eff} = g_J\mu_B\,[J(J+1)]^{0.5}$ = 3.58 $\mu_B$ and 3.62 $\mu_B$ for Pr and Nd, respectively. The paramagnetic Curie temperatures are $\theta_p$ = –19 K for $Pr_3Ni_2B_2N_{3-\delta}$ and $\theta_p$ = +6 K for $Nd_3Ni_2B_2N_{3-\delta}$.

For $Nd_3Ni_2B_2N_{3-\delta}$, low temperature susceptibility, isothermal magnetisation and specific heat results in Fig. 3b, 4a and 4c reveal long range magnetic order below $T_C$ =17 K. The magnetization hardly exceeds 1.2 $\mu_B$/Nd at 6 T, thus referring to a ferrimagnetic ordering of the Nd moments situated in two inequivalent lattice sites. A spin reorientation transition towards a nearly antiferromagnetic state with a small ferrimagnetic component is observed at 10 K, below which the isothermal magnetisation, M (H) shown in Fig. 4a, displays a metamagnetic spin flip transition connected with a strongly hysteretic behaviour and a saturation magnetisation M (3K, 6T)~1.3 $\mu_B$/Nd.

In case of $Pr_3Ni_2B_2N_{3-\delta}$ temperature dependent susceptibility, isothermal magnetisation, and specific heat data in Fig. 3b, Fig. 4b and 4d, respectively, do not reveal any indication for intrinsic long range magnetic order at least down to 2 K. The significant deviation of the susceptibility from a simple Curie-Weiss behaviour below 50 K is due to crystalline electric field effects.

## Summary

Novel quaternary boronitride compounds $Pr_3Ni_2B_2N_{3-\delta}$ and $Nd_3Ni_2B_2N_{3-\delta}$ crystallize in the $La_3Ni_2B_2N_3$ structure-type. AG-type magnetic pair-breaking in solid solutions $La_{3-x}R_xNi_2B_2N_{3-\delta}$ leads to the suppression of superconductivity at critical *R* concentrations $x_{crit.}$~1.9 and 1.0 for Pr and Nd, respectively. For $Pr_3Ni_2B_2N_{3-\delta}$ no long range magnetic order is observed down to 2 K, whereas $Nd_3Ni_2B_2N_{3-\delta}$ shows ferrimagnetic order below $T_C$ = 17 K and a spin reorientation transition towards a nearly antiferromagnetic state at about 10 K.

## Acknowledgement

The research by T.A. was supported by the Higher Education Commision of Pakistan (HEC) under the scholarship scheme in Natural & Basic Sciences from Austria.